\documentclass[12pt]{article}
\usepackage{amssymb,amsmath,epsfig}
\allowdisplaybreaks
\begin{document}
\title{\bf Exact Solutions and Conserved Quantities in $f(R,T)$ Gravity}

\author{M. Sharif \thanks{msharif.math@pu.edu.pk} and Iqra Nawazish
\thanks{iqranawazish07@gmail.com}\\
Department of Mathematics, University of the Punjab,\\
Quaid-e-Azam Campus, Lahore-54590, Pakistan.}

\date{}

\maketitle
\begin{abstract}
This paper explores Noether and Noether gauge symmetries of
anisotropic universe model in $f(R,T)$ gravity. We consider two
particular models of this gravity and evaluate their symmetry
generators as well as associated conserved quantities. We also find
exact solution by using cyclic variable and investigate its behavior
via cosmological parameters. The behavior of cosmological parameters
turns out to be consistent with recent observations which indicates
accelerated expansion of the universe. Next we study Noether gauge
symmetry and corresponding conserved quantities for both isotropic
and anisotropic universe models. We conclude that symmetry
generators and the associated conserved quantities appear in all
cases.
\end{abstract}
{\bf Keywords:} Noether symmetry; Conserved quantity; $f(R,T)$
gravity.\\
{\bf PACS:} 04.20.Jb; 04.50.Kd; 95.36.+x.

\section{Introduction}

In the last century, the crucial observational discoveries
established revolutionary advancements in modern cosmology that
introduced a new vision of the current accelerated expanding
universe. The accelerated epoch of the universe known as ``dark
energy'' (DE) possesses a huge amount of negative pressure. At
theoretical level, the conclusive evidences about accelerated
expansion of the universe and enigmatic behavior of DE lead to
introduce modified theories of gravity. The $f(R)$ gravity is the
simplest proposal ($R$ represents Ricci scalar) developed by
replacing $R$ with a generic function independent of any non-minimal
curvature and matter coupling in the Einstein-Hilbert action.

Different researchers established basic review of $f(R)$ gravity
\cite{2} and also discussed stability of its different models
\cite{3}. The idea of coupling between curvature and matter was
initially presented by Nojiri and Odintsov \cite{4} who explored
explicit and implicit couplings in $f(R)$ gravity. Harko et al.
\cite{5} developed a gravitational theory involving both curvature
as well as matter components known as $f(R,T)$ gravity ($T$ denotes
trace of the energy-momentum tensor). Sharif and Zubair \cite{10}
discussed universe evolution via energy conditions along with
stability criteria, reconstructed different DE models, exact
solutions of anisotropic universe and thermodynamical picture in
$f(R,T)$ gravity.

The discovery of CMBR reveals that the early universe was spatially
homogeneous but largely anisotropic while this anisotropy still
exists in terms of CMB temperature in the present universe. We
consider Bianchi type models which measure the effect of anisotropy
in the early universe through current observations \cite{6}. The
anisotropic universe model indicates that the initial anisotropy
determines the fate of rapid expansion of the early universe which
will continue for initially large values of anisotropy. If the
initial anisotropy is small then the rapid expansion will end
leading to a highly isotropic universe \cite{7}. Akarsu and Kilinc
\cite{8} studied Bianchi type I (BI) model that corresponds to de
Sitter universe for different equation of state (EoS) models. Sharif
and Zubair \cite{9} formulated exact solutions of BI universe model
for power-law and exponential expansions in $f(R,T)$ gravity. Shamir
\cite{a5} discussed exact solutions of locally rotationally
symmetric (LRS) BI universe model and investigated physical behavior
of cosmological parameters in this gravity. Kanakavalli and Ananda
\cite{i} obtained exact solutions of LRS BI model in the presence of
cosmic string source and curvature-matter coupling.

Symmetry approximation plays a crucial role to determine exact
solutions or elegantly reduces complexity of a non-linear system of
equations. Noether symmetry is a useful approach to evaluate unknown
parameters of differential equations. Sharif and Waheed explored
Bardeen model \cite{12} as well as stringy charged black holes
\cite{13} via approximate symmetry. They also evaluated Noether
symmetries of FRW and LRS BI models by including an inverse
curvature term in the action of Brans-Dicke theory \cite{14}.
Kucukakca et al. \cite{19} established exact solutions of LRS BI
universe model through Noether symmetry approach in the same
gravity. Jamil et al. \cite{11} discussed Noether symmetry in
$f(\mathcal{T})$ gravity ($\mathcal{T}$ denotes torsion) that
involves matter as well as scalar field contributions and determined
explicit form of $f(\mathcal{T})$ for quintessence and phantom
models. Kucukakca \cite{16} found exact solutions of flat FRW
universe model via Noether symmetry in scalar-tensor theory
incorporating non-minimal coupling with torsion scalar. Sharif and
Shafique \cite{a1} discussed Noether and Noether gauge symmetries in
this gravity. Sharif and Fatima \cite{a2} explored Noether symmetry
of flat FRW model through vacuum and non-vacuum cases in $f(G)$
gravity.

Capozziello et al. \cite{20} explored Noether symmetry to determine
exact solutions of spherically symmetric spacetime in $f(R)$
gravity. Vakili \cite{21} obtained Noether symmetry of flat FRW
metric and analyzed the behavior of effective EoS parameter in
quintessence phase. Jamil et al. \cite{22} studied Noether symmetry
of flat FRW universe using tachyon model in this gravity. Hussain et
al. \cite{23} studied Noether gauge symmetry of flat FRW universe
model for $f(R)$ power-law model which generates zero gauge term.
Shamir et al. \cite{24} analyzed Noether gauge symmetry for the same
model as well as for static spherically symmetric spacetime and
found non-zero gauge term. Kucukakca and Camci \cite{25} established
Noether gauge symmetry of FRW universe model in Palatini formalism
of $f(R)$ gravity. Momeni et al. \cite{a3} investigated the
existence of Noether symmetry and discussed stability of solutions
for flat FRW universe model in $f(R,T)$ and mimetic $f(R)$ gravity.
They also explored a class of solutions with future singularities.

In this paper, we discuss Noether and Noether gauge symmetries of BI
universe model in $f(R,T)$ gravity. We formulate exact solution of
the field equations to discuss cosmic evolution via cosmological
parameters. The format of this paper is as follows. In section
\textbf{2}, we discuss a basic formalism of $f(R,T)$ gravity,
Noether and Noether gauge symmetries. Section \textbf{3} explores
Noether symmetry of BI model for two theoretical models of $f(R,T)$
gravity and also establish exact solution via cyclic variables. In
section \textbf{4}, we obtain symmetry generator and corresponding
conserved quantities through Noether gauge symmetry for flat FRW and
BI models. In the last section, we summarize the results.

\section{Basic Framework}

The current cosmic expansion successfully discusses not only from
the contribution of the scalar-curvature part but also describes
from a non-minimal coupling between curvature and matter components
as well. This non-minimal coupling yields non-zero divergence of the
energy-momentum tensor due to which an extra force appears that
deviates massive test particles from geodesic trajectories. The
action of such modified gravity is given by \cite{5}
\begin{equation}\label{1}
\mathcal{I}=\int d^4x\sqrt{-g}[\frac{f(R,T)}{2\kappa^2}+
\mathcal{L}_m],
\end{equation}
where $f$ describes a simple coupling of geometry and matter whereas
$\mathcal{L}_m$ denotes the matter Lagrangian. The variation of
action (\ref{1}) with respect to $g_{\mu\nu}$ yields non-linear
partial differential equation of the following form
\begin{eqnarray}\nonumber
&&f_R(R,T)
R_{\mu\nu}-\frac{1}{2}f(R,T)g_{\mu\nu}+(g_{\mu\nu}\Box-\nabla_\mu\nabla_\nu)
f_R(R,T)+f_T(R,T)T_{\mu\nu}\\\label{2}&&+f_T(R,T)\Theta_{\mu\nu}=\kappa^2T_{\mu\nu},
\end{eqnarray}
where $\nabla_{\mu}$ shows covariant derivative and
\begin{eqnarray}\nonumber
\Box&=&\nabla_{\mu}\nabla^{\mu},\quad f_R(R,T)=\frac{\partial
f(R,T)}{\partial R},\quad f_T(R,T)=\frac{\partial f(R,T)}{\partial
T},\\\nonumber\Theta_{\mu\nu}&=&\frac{g^{\alpha\beta}\delta
T_{\alpha\beta}}{\delta
g^{\mu\nu}}=g_{\mu\nu}\mathcal{L}_m-2T_{\mu\nu}-2g^{\alpha\beta}
\frac{\partial^2\mathcal{L}_m}{\partial g^{\alpha\beta}\partial
g^{\mu\nu}}.
\end{eqnarray}
The trace of Eq.(\ref{2}) provides a significant relationship
between geometric and matter parts as follows
\begin{equation}\nonumber
Rf_R(R,T)+3\Box f_R(R,T)-2f(R,T)+Tf_T(R,T)+\Theta
f_T(R,T)=\kappa^2T.
\end{equation}
Harko et al \cite{5} introduced some theoretical models for
different choices of matter as
\begin{itemize}
\item $f(R,T)$=$R+2f(T)$,
\item $f(R,T)$=$f_1(R)+f_2(T)$,
\item $f(R,T)$=$f_1(R)+f_2(R)f_3(T)$.
\end{itemize}

Noether symmetry is the most significant approach to deal with
non-linear partial differential equations. The existence of Noether
symmetry is possible only if Lie derivative of Lagrangian vanishes,
i.e., the vector field is unique on the tangent space. In such
situation, the vector field behaves as a symmetry generator which
further generates conserved quantity. Noether gauge symmetry being
generaliztion of Noether symmetry preserves some extra symmetries
along a non-vanishing gauge term. The vector field and its first
order prolongation are defined as
\begin{eqnarray}\nonumber
K&=&\xi(t,q^i)\frac{\partial}{\partial
t}+\eta^j(t,q^i)\frac{\partial}{\partial q^j},\\\nonumber
K^{[1]}&=&K+(\eta^j,_t+\eta^j,_i\dot{q}^i-\xi,_t\dot{q}^j
-\xi,_i\dot{q}^i\dot{q}^j)\frac{\partial}{\partial\dot{q}^i},
\end{eqnarray}
where $t$ identifies as affine parameter, $\xi,~\eta$ are symmetry
generator coefficients, $q^i$ represents $n$ generalized positions
and dot denotes time derivative. The vector field $K$ generates
Noether gauge symmetry if Lagrangian preserves the following
condition
\begin{equation}\nonumber
K^{[1]}\mathcal{L}+(D\xi)\mathcal{L}=DG(t,q^i).
\end{equation}
Here $G(t,q^i)$ represents the gauge term and $D$ denotes the total
derivative operator defined as
\begin{eqnarray}\nonumber
D&=&\frac{\partial}{\partial t}+\dot{q}^i\frac{\partial}{\partial
q^i}.
\end{eqnarray}
According to Noether theorem, there exists a conserved quantity
corresponding to each symmetry of a system. In case of Noether gauge
symmetry, the conserved quantity for vector field $K$ takes the form
\begin{equation}\nonumber
\Sigma=G-\xi\mathcal{L}-(\eta^j-\dot{q}^j\xi)
\frac{\partial\mathcal{L}}{\partial\dot{q}^j}.
\end{equation}
For the existence of Noether symmetry, the following condition must
holds
\begin{equation*}
L_K\mathcal{L}=K\mathcal{L}=0,
\end{equation*}
where $L$ represents Lie derivative while the vector field $K$ and
conserved quantity corresponding to symmetry generator turn out to
be
\begin{equation}\label{a}
K=\beta^i(q^i)\frac{\partial}{\partial
q^i}+\left[\frac{d}{dt}(\beta^i(q^i))\right]\frac{\partial}{\partial\dot{q}^i},
\quad \Sigma=-\eta^j\frac{\partial\mathcal{L}}{\partial\dot{q}^j}.
\end{equation}
The equation of motion and associated Hamiltonian equation of a
dynamical system are defined as
\begin{eqnarray}\nonumber
\frac{\partial\mathcal{L}}{\partial
q^i}-\frac{d}{dt}\left(\frac{\partial\mathcal{L}}{\partial
\dot{q}^i}\right)=0,\quad
\Sigma_i\dot{q}^ip_i-\mathcal{L}=\mathcal{H},\quad
p_i=\frac{\partial\mathcal{L}}{\partial q^i},
\end{eqnarray}
where $p_i$ represents conjugate momenta of configuration space.

\section{Noether Symmetry for BI Universe Model}

Here we apply Noether symmetry approach to deal with non-linear
partial differential equation (\ref{2}) and evaluate symmetry
generators as well as corresponding conserved quantities of BI
universe model given by
\begin{equation}\label{3}
ds^2=-dt^2+a^2(t)dx^2+b^2(t)(dy^2+dz^2),
\end{equation}
where $t$ denotes cosmic time, scale factors $a$ and $b$ measure
expansion of the universe in $x$ and $y,~z$-directions,
respectively. We consider the perfect fluid distribution given by
\begin{equation*}
T_{\mu\nu}=(\rho+p)u_\mu u_\nu+pg_{\mu\nu},
\end{equation*}
where $p,~\rho$ and $u_\mu$ represent pressure, energy density and
four-velocity of the fluid, respectively. To evaluate the
Lagrangian, we rewrite the action (\ref{1}) as
\begin{equation}\label{4}
\mathcal{I}=\int\sqrt{-g}[f(R,T)-\lambda(R-\bar{R})-\chi(T-\bar{T})
+\mathcal{L}_m]dt,
\end{equation}
where $\sqrt{-g}=ab^2$, $\bar{R},~\bar{T}$ are dynamical constraints
while $\lambda,~\chi$ are Lagrange multipliers given by
\begin{eqnarray}\nonumber
&&\bar{R}=\frac{2}{ab^2}(\ddot{a}b^2+2ab\ddot{b}
+2b\dot{a}\dot{b}+a\dot{b^2}),\quad
\bar{T}=3p(a,b)-\rho(a,b),\\\nonumber &&\lambda=f_R(R,T),\quad
\chi=f_T(R,T).
\end{eqnarray}
The field equation (\ref{2}) is not easy to tackle with perfect
fluid configuration and also there is no unique definition of matter
Lagrangian. In order to construct Lagrangian, we consider
$\mathcal{L}_m=p(a,b)$ which yields
\begin{eqnarray}\nonumber
&&\mathcal{L}(a,b,R,T,\dot{a},\dot{b},\dot{R},\dot{T})=ab^2[f(R,T)-Rf_R(R,T)
-Tf_T(R,T)\\\nonumber&&+f_T(R,T)(3p(a,b)-\rho(a,b))+p(a,b)]
-(4b\dot{a}\dot{b}+2a\dot{b}^2)f_R(R,T)
\\\label{5}&&-(2b^2\dot{a}\dot{R}+4ab\dot{b}\dot{R})
f_{RR}(R,T)-(2b^2\dot{a}\dot{T}+4ab\dot{b}\dot{T})f_{RT}(R,T).
\end{eqnarray}

The corresponding equations of motion and energy function of
dynamical system become
\begin{eqnarray}\nonumber
&&\frac{\dot{b^2}}{b^2}+\frac{2\ddot{b}}{b}=-\frac{1}{2f_R(R,T)}
[f(R,T)-Rf_R(R,T)-Tf_T(R,T)+f_T(R,T)\\\nonumber&&
\times(3p(a,b)-\rho(a,b))+p(a,b)
+a\{f_T(3p,_{_a}-\rho,_{_a})+p,_{_a}\}+\frac{4\dot{b}\dot{R}f_{RR}(R,T)}
{b}\\\nonumber&&+2\ddot{R}f_{RR}(R,T)+2\dot{R}^2f_{RRR}(R,T)
+4\dot{R}\dot{T}f_{RRT}(R,T)+2\ddot{T}f_{RT}(R,T)\\\label{24}
&&+2\dot{T}^2f_{RTT}(R,T)],
\\\nonumber&&\frac{\ddot{a}}{a}+\frac{\dot{a}\dot{b}}{ab}+\frac{\ddot{b}}{b}
=-\frac{1}{4f_R(R,T)}[2(f(R,T)-Rf_R(R,T)-Tf_T(R,T)\\\nonumber&&+f_T(R,T)
(3p(a,b)-\rho(a,b))+p(a,b))+b\{f_T(3p,_{_b}-\rho,_{_b})+p,_{_b}\}]\\\nonumber&&+2(a^{-1}\dot{a}
\dot{R}+\ddot{R})f_{RR}+2\dot{R}^2f_{RRR}+2(a^{-1}\dot{a}
\dot{T}+\ddot{T})f_{RT}+2(b^{-1}\dot{b}\dot{R}+2\dot{R}\dot{T}\\\label{25}&&+\dot{T}^2)f_{RRT}
+2b^{-1}\dot{b}\dot{T}f_{RTT}=0,
\\\nonumber&&\frac{\dot{b^2}}{b^2}+\frac{2\dot{a}\dot{b}}{ab}=-\frac{1}{f_R(R,T)}
\left[\left(\frac{2\dot{b}\dot{R}}{b}+\frac{\dot{a}\dot{R}}{a}\right)f_{RR}(R,T)
+\left(\frac{2\dot{b}\dot{T}}{b}+\frac{\dot{a}\dot{T}}{a}\right)\right.
\\\nonumber&&\times\left.f_{RT}(R,T)
+\frac{1}{2}(f(R,T)-Rf_R(R,T)-Tf_T(R,T)+f_T(R,T)(3p(a,b)\right.
\\\label{26}&&\left.-\rho(a,b))+p(a,b))\right].
\end{eqnarray}
The conjugate momenta corresponding to configuration space
($a,~b,~R,~T$) are
\begin{eqnarray}\label{9}
p_a&=&\frac{\partial\mathcal{L}}{\partial\dot{a}}=-4b\dot{b}f_R(R,T)
-2b^2(\dot{R}f_{RR}(R,T)+\dot{T}f_{RT}(R,T)),\\\label{10}
p_b&=&\frac{\partial\mathcal{L}}{\partial\dot{b}}=-4f_R(R,T)(a\dot{b}+b\dot{a}
-4ab(\dot{R}f_{RR}(R,T)+\dot{T}f_{RT}(R,T)),\\\label{11}p_R&=&\frac{\partial\mathcal{L}}
{\partial\dot{R}}= -(4ab\dot{b}+2b^2\dot{a})f_{RR}(R,T),\\\label{12}
p_T&=&\frac{\partial\mathcal{L}} {\partial\dot{T}}=
-(4ab\dot{b}+2b^2\dot{a})f_{RT}(R,T).
\end{eqnarray}
For Noether symmetry, the vector field (\ref{a}) takes the following
form
\begin{equation}\label{6}
K=\alpha\frac{\partial}{\partial a}+\beta\frac{\partial}{\partial
b}+\gamma\frac{\partial}{\partial R}+\delta\frac{\partial}{\partial
T}+\dot{\alpha}\frac{\partial}{\partial
\dot{a}}+\dot{\beta}\frac{\partial}{\partial
\dot{b}}+\dot{\gamma}\frac{\partial}{\partial
\dot{R}}+\dot{\delta}\frac{\partial}{\partial \dot{T}},
\end{equation}
where $\alpha,~\beta,~\gamma$ and $\delta$ are unknown coefficients
of generator which depend on variables $a,~b,~R$ and $T$ while the
time derivatives of these coefficients are
\begin{eqnarray}\nonumber
\dot{\alpha}&=&\dot{a}\frac{\partial\alpha}{\partial
a}+\dot{b}\frac{\partial\alpha}{\partial
b}+\dot{R}\frac{\partial\alpha}{\partial
R}+\dot{T}\frac{\partial\alpha}{\partial T},\quad
\dot{\beta}=\dot{a}\frac{\partial\beta}{\partial
a}+\dot{b}\frac{\partial\beta}{\partial
b}+\dot{R}\frac{\partial\beta}{\partial
R}+\dot{T}\frac{\partial\beta}{\partial
T},\\\label{7}\\\label{8}\dot{\gamma}&=&\dot{a}\frac{\partial\gamma}{\partial
a}+\dot{b}\frac{\partial\gamma}{\partial
b}+\dot{R}\frac{\partial\gamma}{\partial
R}+\dot{T}\frac{\partial\gamma}{\partial
T},\quad\dot{\delta}=\dot{a}\frac{\partial\delta}{\partial
a}+\dot{b}\frac{\partial\delta}{\partial
b}+\dot{R}\frac{\partial\delta}{\partial
R}+\dot{T}\frac{\partial\delta}{\partial T}.
\end{eqnarray}
Taking Lie derivative of Lagrangian (\ref{5}) for vector field
(\ref{6}) and inserting Eqs.(\ref{7}) and (\ref{8}), we obtain the
following over determined system of equations by comparing the
coefficients of
$\dot{a}^2,~\dot{b}^2,~\dot{R}^2,~\dot{T}^2,~\dot{a}\dot{b},~\dot{a}
\dot{R},~\dot{a}\dot{T},~\dot{b}\dot{R},~\dot{b}\dot{T}$ and
$\dot{R}\dot{T}$ as
\begin{eqnarray}\label{13}
&&(b\alpha,_{_R}+2a\beta,_{_R})f_{RR}=0,\\\label{14}
&&(b\alpha,_{_T}+2a\beta,_{_T})f_{RT}=0,\\\label{15}&&2\beta,_{_a}f_R
+b\gamma,_{_a}f_{RR}+b\delta,_{_a}f_{RT}=0,\\\label{16}&&b\alpha,_{_R}f_{RT}
+b\alpha,_{_T}f_{RR}+2a\beta,_{_R}f_{RT}+2a\beta,_{_T}f_{RR}=0,\\\nonumber&&2\beta
f_{RR}+b\gamma f_{RRR}+b\delta
f_{RRT}+b\alpha,_{_a}f_{RR}+2a\beta,_{_a}f_{RR}+2\beta,_{_R}f_R
+b\gamma,_{_R}f_{RR}\\\label{18}&&+b\delta,_{_R}f_{RT}=0,\\\nonumber&&2\beta
f_{RT}+b\gamma f_{RRT}+b\delta
f_{RTT}+b\alpha,_{_a}f_{RT}+2a\beta,_{_a}f_{RT}+2\beta,_{_T}f_R
+b\gamma,_{_T}f_{RR}\\\label{19}&&+b\delta,_{_T}f_{RT}=0,\\\nonumber&&2\beta
f_R+2b\gamma f_{RR}+2b\delta
f_{RT}+2b\alpha,_{_a}f_R+2a\beta,_{_a}f_R+2b\beta,_{_b}f_R+2ab
\gamma,_{_a}f_{RR}\\\label{20}&&+b^2\gamma,_{_b}f_{RR}+2ab\delta,_{_a}f_{RT}
+b^2\delta,_{_b}f_{RT}=0,\\\nonumber&&2b\alpha f_{RR}+2a\beta
f_{RR}+2ab\gamma f_{RRR}+2ab\delta
f_{RRT}+b^2\alpha,_{_b}f_{RR}+2b\alpha,_{_R}f_R+2ab\\\label{21}&&\times\beta,_{_b}f_{RR}
+2a\beta,_{_R}f_R+2ab\gamma,_{_R}f_{RR}+2ab\delta,_{_R}f_{RT}=0,\\\nonumber&&2b\alpha
f_{RT}+2a\beta f_{RT}+2ab\gamma f_{RRT}+2ab\delta
f_{RTT}+b^2\alpha,_{_b}f_{RT}+2b\alpha,_{_T}f_R+2ab\\\label{22}&&\times\beta,_{_b}f_{RT}
+2a\beta,_{_T}f_R+2ab\gamma,_{_T}f_{RR}+2ab\delta,_{_T}f_{RT}=0,\\\nonumber
&&\alpha f_R+a\gamma f_{RR}+a\delta
f_{RT}+2b\alpha,_{_b}f_R+2a\beta,_{_b}f_R+2ab\gamma,_{_b}f_{RR}
+2ab\delta,_{_b}\\\label{17}&&\times f_{RT}=0,
\\\nonumber
&&b^2\alpha[f-Rf_R-Tf_T+f_T(3p-\rho)+p+a\{f_T(3p,_{_a}-\rho,_{_a})+p,_{_a}\}]
+\beta[2ab\\\nonumber&&\times(f-Rf_R-Tf_T+f_T(3p-\rho)+p)
+ab^2\{f_T(3p,_{_b}-\rho,_{_b})+p,_{_b}\}]
+ab^2\\\nonumber&&\times\gamma[-(Rf_{RR}+Tf_{RT})+f_{RT}(3p-\rho)]
+ab^2\delta[-(Rf_{RT}+Tf_{TT})+f_{TT}\\\label{23}&&\times(3p-\rho)]=0.
\end{eqnarray}
We solve this non-linear system of partial differential equations
for two models of $f(R,T)$ gravity and evaluate possible solutions
of symmetry generator coefficients as well as corresponding
conserved quantities.

\subsection{\textbf{$f(R,T)=R+2f(T)$}}

Here we discuss a solution for a simple model that explores Einstein
gravity with matter components such as $f(R,T)=R+2f(T)$, where the
curvature term behaves as a leading term of the model. This model
corresponds to $\Lambda$CDM model when matter part comprises
cosmological constant as a function of trace $T$. Consequently, this
model reduces to
\begin{equation}\label{27}
f(R,T)=R+2\Lambda+h(T).
\end{equation}
To find the solution of Eqs.(\ref{13})-(\ref{23}), we consider
power-law form of unknown coefficients of vector field as
\begin{eqnarray}\label{28}
\alpha&=&\alpha_0a^{\alpha_1}b^{\alpha_2}R^{\alpha_3}T^{\alpha_4},
\quad\beta=\beta_0a^{\beta_1}b^{\beta_2}R^{\beta_3}T^{\beta_4},
\\\label{29}\gamma&=&\gamma_0a^{\gamma_1}b^{\gamma_2}R^{\gamma_3}T^{\gamma_4},
\quad\delta=\delta_0a^{\delta_1}b^{\delta_2}R^{\delta_3}T^{\delta_4},
\end{eqnarray}
where the powers are unknown constants to be determined. Using these
coefficients in Eqs.(\ref{13})-(\ref{22}), we obtain
\begin{eqnarray*}
\alpha_0&=&-\beta_0(\alpha_2+2),\quad\alpha_1=1,\quad\alpha_3=0,
\quad\alpha_4=0,\quad\gamma=0,
\\\nonumber\beta_1&=&0,\quad\beta_2=\alpha_2+1,\quad\beta_3=0,\quad\beta_4=0.
\end{eqnarray*}
Inserting these values in Eq.(\ref{28}), it follows that
\begin{equation*}
\alpha=-\beta_0(\alpha_2+2)ab^{\alpha_2},\quad\beta=\beta_0b^{\alpha_2+1}.
\end{equation*}
In order to evaluate $\alpha_2$, we substitute these solutions in
Eq.(\ref{17}) which implies that either $\alpha_2=0$ or
$\alpha_2=\frac{1}{2}$.

\subsubsection*{Case I: $\alpha_2=0$}

In this case, the generator coefficients turn out to be
\begin{equation*}
\alpha=-2\beta_0a,\quad\beta=\beta_0b.
\end{equation*}
In order to evaluate the remaining coefficients, we insert these
values in Eqs.(\ref{24}), (\ref{26}) and (\ref{23}) which give
\begin{eqnarray*}
h(T)&=&l_1T+l_2,\quad\delta=0,\quad
p=l_3a^{-\frac{1}{5}}b^{-\frac{2}{5}},\\
\rho&=&-\frac{1}{2l_1}[2\Lambda+l_2+(3l_1-1)
l_3a^{-\frac{1}{5}}b^{-\frac{2}{5}}].
\end{eqnarray*}
Substituting all these solutions in Eqs.(\ref{13})-(\ref{22}), we
obtain $l_1=-\frac{19}{3}$. Consequently, the coefficients of
symmetry generator and $f(R,T)$ model become
\begin{equation*}
\alpha=-2\beta_0a,\quad\beta=\beta_0b,\quad\gamma=0,\quad\delta=0,\quad
f(R,T)=R-\frac{19T}{3},
\end{equation*}
where $h(T)=-\frac{19T}{3}-2\Lambda$ and
$T=\frac{87}{19}l_3a^{-\frac{1}{5}}b^{-\frac{2}{5}}$. To avoid
Dolgov-Kawasaki instability, the $f(R,T)$ model preserves the
following conditions \cite{r4}
\begin{equation}\label{R}
f_R(R)>0,\quad f_{RR}(R)>0,\quad 1+f_T(R,T)>0,\quad R>R_0.
\end{equation}
In this case, the constructed $f(R,T)$ model is found to be viable
for $l_3<0$. Using the values of symmetry generator coefficients, we
obtain symmetry generator which yields scaling symmetry and its
conserved quantity as
\begin{equation*}
K=-2\beta_0a\frac{\partial}{\partial
a}+\beta_0b\frac{\partial}{\partial
b},\quad\Sigma=\beta_0[-4ab\dot{b}+4\dot{a}b^{2}].
\end{equation*}

Now we solve the field equations using cyclic variable whose
existence is assured by the presence of symmetry generator of
Noether symmetry. We consider a point transformation which reduces
complex nature of the system to $\phi: (a,b)\rightarrow(v,z)$
implying that $\phi_Kdv=0$ and $\phi_Kdz=1$. The second mapping
indicates that the Lagrangian must be free from the variable $z$.
Imposing this point transformation, we reduce the complexity of the
system as
\begin{equation}\label{30}
v=\zeta_0a^{\frac{1}{2}}b,\quad z=\frac{\ln b}{\beta_0},
\end{equation}
where $z$ is cyclic variable and $\zeta_0$ denotes arbitrary
constant. The inverse point transformation of variables yields
\begin{equation}\label{31}
a=\zeta_1v^{\frac{1}{2}}e^{-2\beta_0z},\quad
b=\zeta_2e^{\beta_0z},\quad\rho=-\frac{30\zeta_3v^{-\frac{2}{5}}}{19},\quad
p=\zeta_3v^{-\frac{2}{5}}.
\end{equation}
Here we redefine arbitrary constants as
$\zeta_3=l_3\zeta_1^{-\frac{1}{5}}\zeta_2^{-\frac{2}{5}}$. For the
above solutions, the Lagrangian (\ref{5}) and the corresponding
equations of motion with associated energy function
(\ref{24})-(\ref{26}) take the form
\begin{eqnarray*}
&&\mathcal{L}=\zeta_4(4\beta_0v^{\frac{-1}{2}}\dot{v}\dot{z}+4\beta_0^2
v^{\frac{1}{2}}\dot{z}^2-30v^{\frac{2}{5}}),\\\nonumber
&&2\beta_0v^{\frac{-1}{2}}\ddot{z}+2\beta_0^2v^{-\frac{1}{2}}\dot{z}^2
-12v^{-\frac{3}{5}}=0,
\\\nonumber&&8\beta_0v^{\frac{1}{2}}\ddot{z}+v^{-\frac{3}{2}}\dot{v}^2
+4\beta_0v^{-\frac{1}{2}}\dot{z}-2v^{-\frac{1}{2}}\ddot{v}=0,
\\\nonumber&&30v^{\frac{2}{5}}+4\beta_0^2v^{\frac{1}{2}}\dot{z}^2
+\beta_0v^{-\frac{3}{2}}\dot{v}^2\dot{z}-2\beta_0v^{-\frac{1}{2}}\dot{v}\ddot{z}=0.
\end{eqnarray*}
We solve the above equations to evaluate the time dependent
solutions of new variables ($v,~z$)
\begin{equation*}
v=2(t-\zeta_4)^{\frac{1}{2}}(t^2-2t+\zeta_4^2),\quad
z=\frac{1}{12\beta_0}[12\beta_0\zeta_5-2.93
-4\ln[(t-\zeta_4)^{\frac{5}{2}}]],
\end{equation*}
where $\zeta_4$ and $\zeta_5$ represent integration constants.
Inserting these values into Eq.(\ref{31}), we obtain
\begin{eqnarray}\label{33}
a&=&\frac{8}{5}\zeta_1e^{-2\beta_0\zeta_5}(t-\zeta_4)^{\frac{5}{3}},
\quad
b=\frac{8}{5}\zeta_2e^{\beta_0\zeta_5}(t-\zeta_4)^{-\frac{1}{3}}
(t^2-2t\zeta_1+\zeta_1^2),\\\nonumber\rho&=&-\frac{30\zeta_3}{19}
[2(t-\zeta_4)^{\frac{1}{2}}(t^2-2t+\zeta_4^2)]^{-\frac{2}{5}}, \quad
p=\zeta_3[2(t-\zeta_4)^{\frac{1}{2}}(t^2-2t+\zeta_4^2)]^{-\frac{2}{5}}.\\\label{32}
\end{eqnarray}

We study the behavior of some well-known cosmological parameters
like Hubble, deceleration and EoS parameters using scale factors and
matter contents. These parameters play significant role to discuss
cosmic expansion as Hubble parameter $(H)$ measures the rate of
cosmic expansion while deceleration parameter $(q)$ determines that
either expansion is accelerated $(q<0)$ or decelerated $(q>0)$ or
constant expansion ($q=0$). The EoS parameter
$(\omega=\frac{p}{\rho})$ evaluates different eras of the universe
and also differentiates DE era into different phases like
quintessence ($-1<\omega\leq-1/3$) or phantom ($\omega<-1$). In case
of BI universe model, the Hubble and deceleration parameters are
\begin{equation}\nonumber
H=\frac{1}{3}\left(\frac{\dot{a}}{a}+\frac{2\dot{b}}{b}\right),\quad
q=-\frac{\dot{H}}{H^2}-1.
\end{equation}
Using Eq.(\ref{33}), the Hubble and deceleration parameters turn out
to be
\begin{equation}\label{a'}
H=\frac{5\zeta_6}{3}\left(1+\frac{t}{\zeta_6}\right),\quad
q=-\frac{3}{5}(\zeta_6+t)^{-2}-1,
\end{equation}
where $\zeta_6=-\zeta_4$. Inserting Eqs.(\ref{33}) and (\ref{32}) in
(\ref{24}) and (\ref{26}), the effective EoS parameter becomes
\begin{equation*}
\omega_{eff}=\frac{p_{eff}}{\rho_{eff}}=1-\frac{\zeta_4-t+3(\sqrt{t-\zeta_4}
(t^2-2t+\zeta_4^2))^{\frac{2}{5}}}{t-\zeta_4}.
\end{equation*}
The crucial pair of ($r,s$) parameters study the correspondence
between constructed and standard universe models such as for
($r,s$)=(1,0), the constructed model corresponds to standard
constant cosmological constant cold dark matter ($\Lambda$CDM)
model. In terms of Hubble and deceleration parameters, these are
defined as
\begin{equation*}
r=q+2q^2-\frac{\dot{q}}{H},\quad s=\frac{r-1}{3(q-\frac{1}{2})}.
\end{equation*}
Using Eq.(\ref{a'}), these parameters take the form
\begin{eqnarray*}
r&=&1+\frac{18}{25}\left(2(t-\zeta_4)^{-4}-2(t-\zeta_4)^{-3}
+(t-\zeta_4)^{-2}\right),\\\nonumber
s&=&\frac{1}{3}(r-1)\left(-\frac{3(t+\zeta_6)^{-2}}{5}-\frac{3}{2}\right)^{-1}.
\end{eqnarray*}
\begin{figure}\epsfig{file=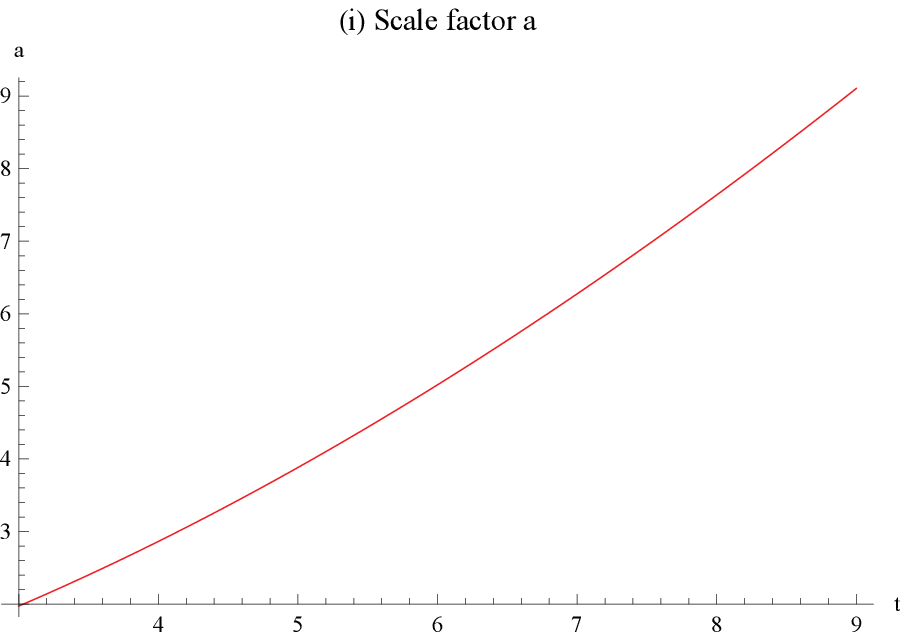,
width=0.45\linewidth}\epsfig{file=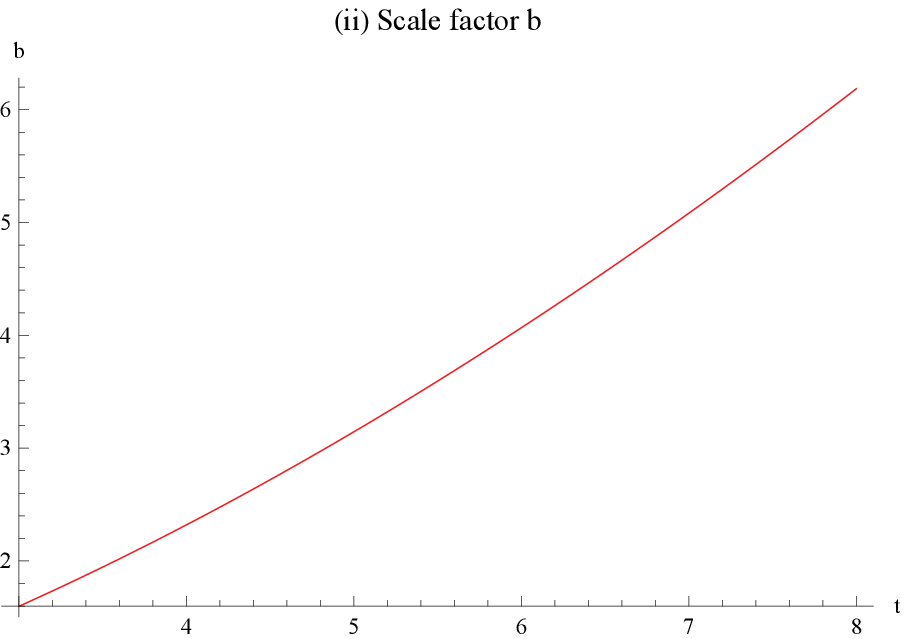,
width=0.45\linewidth}\caption{Plots of scale factors versus cosmic
time $t$: (i) $a(t)$ versus $t$;\newline(ii) $b(t)$ versus $t$ for
$\zeta_1=0.15$, $\zeta_2=0.09$, $\zeta_4=-0.99$, $\zeta_5=0.5$,
$\beta_0=0.1$.}
\end{figure}

Both plots of Figure \textbf{1} represent graphical analysis of the
scale factors $a$ and $b$ which show the increasing behavior of both
scale factors in $x$ and $y,~z$-directions, respectively. This
increasing nature of scale factors indicates the cosmic accelerated
expansion in all directions. The graphical analysis of Hubble and
deceleration parameters is shown in Figure \textbf{2}. Figure
\textbf{2(i)} shows that the Hubble parameter grows continuously
representing expanding universe whereas Figure \textbf{2(ii)} shows
negative deceleration parameter which corresponds to accelerated
expansion of the universe. In Figure \textbf{3}, the first plot
indicates that the effective EoS parameter corresponds to
quintessence phase while Figure \textbf{3(ii)} represents
correspondence of the constructed model with standard $\Lambda$CDM
universe model. Thus, the analysis of cosmological parameters
implies that the universe experiences accelerated expansion for BI
universe model in the context of $f(R,T)$ gravity.
\begin{figure}
\epsfig{file=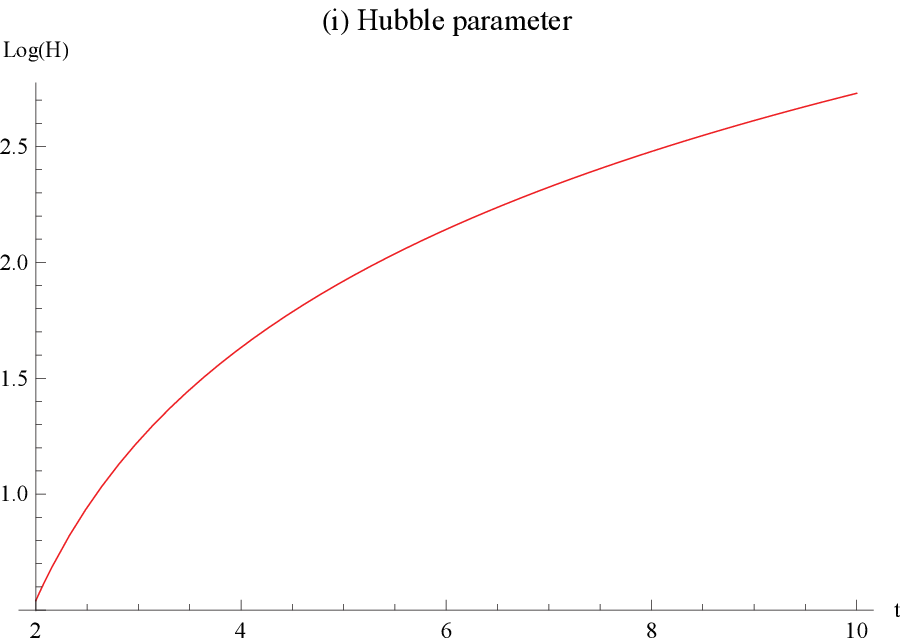, width=0.45\linewidth}\epsfig{file=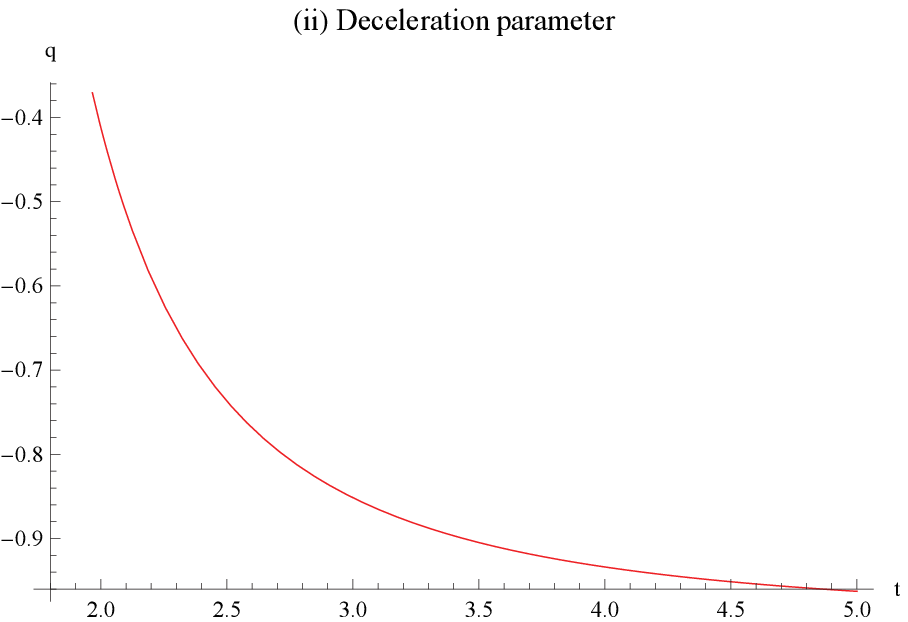,
width=0.45\linewidth}\caption{Plots of (i) Hubble parameter and (ii)
deceleration parameter versus cosmic time $t$ for $\zeta_6=-0.99$.}
\end{figure}
\begin{figure}
\epsfig{file=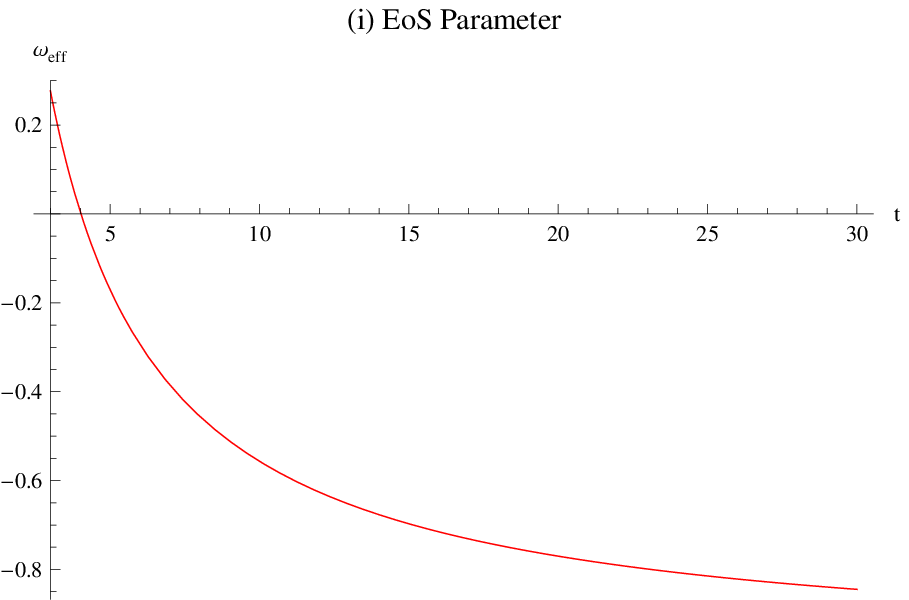, width=0.45\linewidth}\epsfig{file=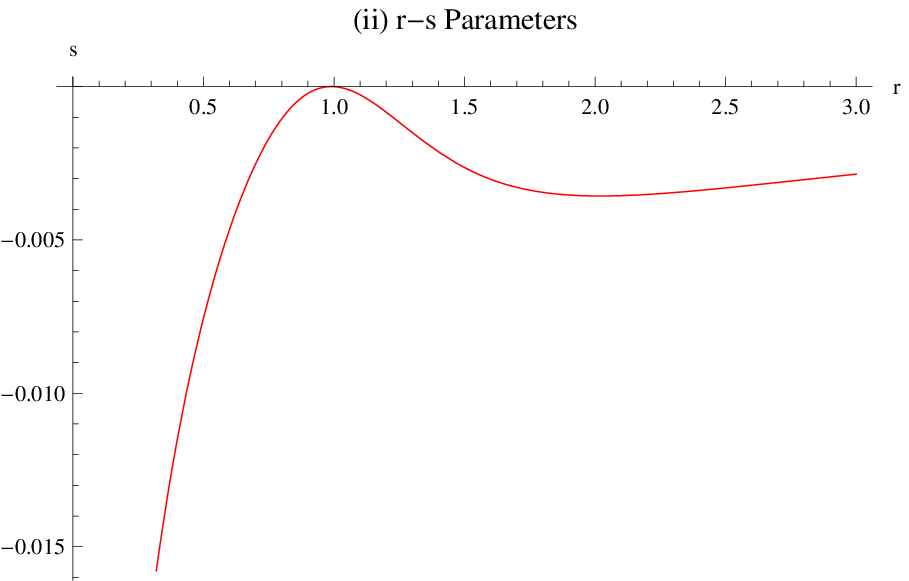,
width=0.45\linewidth}\caption{Plots of (i) EoS Parameter and (ii)
r-s parameters versus cosmic time $t$ for $\zeta_6=-0.99$.}
\end{figure}

\subsubsection*{Case II: $\alpha_2 = \frac{1}{2}$}

For $\alpha_2=\frac{1}{2}$, the solutions become
\begin{equation*}
\alpha=-\frac{5}{2}\beta_0ab^{\frac{1}{2}},\quad\beta=\beta_0b^{\frac{3}{2}},
\end{equation*}
whereas Eq.(\ref{23}) yields
\begin{eqnarray*}
\delta&=&0,\quad h(T)=-2\Lambda+c_1T,\quad
p=c_2a^{\frac{3c_1^2-3c_1-1}{3c_1-1}}b^{\frac{3(5c_1^2-4c_1-2)}{2(3c_1-1)}},
\\\nonumber\rho&=&\left(\frac{3c_1-1}
{c_1-2}\right)c_2a^{\frac{3c_1^2-3c_1-1}{3c_1-1}}b^{\frac{3(5c_1^2-4c_1-2)}{2(3c_1-1)}}.
\end{eqnarray*}
The above solutions satisfy the system of Eqs.(\ref{13})-(\ref{22})
for $c_1=\frac{3\pm\sqrt{21}}{6}$. Under this condition, the
solutions and considered model of $f(R,T)$ gravity take the
following form
\begin{eqnarray*}
\alpha&=&-\frac{5}{2}\beta_0ab^{\frac{1}{2}},\quad\beta=\beta_0b^{\frac{3}{2}},
\quad\gamma,\delta=0,\quad
h(T)=-2\Lambda+\left(\frac{3\pm\sqrt{21}}{6}\right)T,\\\nonumber
p&=&c_2b^{\frac{1}{2}},\quad
\rho=\left(\frac{-3\mp\sqrt{21}}{9\mp\sqrt{21}}\right)c_2b^{\frac{1}{2}},\quad
f(R,T)=R+\left(\frac{3\pm\sqrt{21}}{6}\right)T,
\end{eqnarray*}
where
$T=\left(\frac{30\mp2\sqrt{21}}{9\mp\sqrt{21}}\right)c_2b^{\frac{1}{2}}$.
Here, the constructed model ignores Dolgov-Kawasaki instability as
$f_R,~f_{RR},~1+f_T>0$. The symmetry generator and its corresponding
conserved quantity turn out to be
\begin{equation*}
K=-\frac{5}{2}\beta_0ab^{\frac{1}{2}}\frac{\partial}{\partial
a}+\beta_0b^{\frac{3}{2}}\frac{\partial}{\partial
b},\quad\Sigma=\beta_0[6ab^{\frac{3}{2}}\dot{b}-4\dot{a}b^{\frac{5}{2}}].
\end{equation*}
We consider $z$ to be a cyclic variable which yields
\begin{equation}\nonumber
v=\chi_0a^{\frac{2}{5}}b,\quad z=-\frac{2
b^{-\frac{1}{2}}}{\beta_0},
\end{equation}
where $\chi_0$ denotes arbitrary constant. The corresponding inverse
point transformation leads to
\begin{eqnarray}\nonumber
a&=&\chi_1v^{\frac{5}{2}}\left(-\frac{\beta_0z}{2}\right)^5,\quad
b=\chi_2\left(-\frac{\beta_0z}{2}\right)^{-2},\\\nonumber\quad
p&=&c_2\chi_2\left(-\frac{\beta_0z}{2}\right)^{-1}\quad
\rho=\left(\frac{-3\mp\sqrt{21}}{9\mp\sqrt{21}}\right)c_2\chi_2
\left(-\frac{\beta_0z}{2}\right)^{-1},
\end{eqnarray}
where $\chi_1,\chi_2$ are arbitrary constants. For these solutions,
the Lagrangian (\ref{5}) becomes
\begin{eqnarray*}\nonumber
\mathcal{L}&=&-2\beta_0\chi_1\chi_2^2\left[5v^{\frac{3}{2}}\dot{v}
-6\beta_0v^{\frac{5}{2}}\dot{z}^2\left(-\frac{\beta_0z}{2}\right)^{-1}\right]
+c_2v^{\frac{5}{2}}\left[4\left(\frac{3\pm\sqrt{21}}{6}\right)\right.
\\\nonumber&\times&\left.
\left(\frac{6\mp\sqrt{21}}{9\mp\sqrt{21}}\right)-1\right],
\end{eqnarray*}
which depends upon the cyclic variable $z$. Thus, the resulting
symmetry generator for $\alpha_2=0$ yields scaling symmetry
providing more significant results as compared to
$\alpha_2=\frac{1}{2}$.

\subsection{\textbf{$f(R,T)=f_1(R)+f_2(T)$}}

Here we consider $f(R,T)$ model which does not encourage any
absolute non-minimal coupling of curvature and matter. For vector
field $K$ (\ref{6}), we substitute this model in
Eqs.(\ref{13})-(\ref{20}) and (\ref{22}) yielding the coefficients
of symmetry generator in the form
\begin{eqnarray*}
\alpha&=&-\frac{2ac_3}{b\sqrt{f'_1(R)}}-2ac_4\ln(f'_1(R))-\frac{2c_5}{\sqrt{b}}
-4\ln(b)ac_4-6\ln(b)c_6a+c_7a,
\\\nonumber\beta&=&\frac{c_3}{\sqrt{f'_1(R)}}+(c_8
+\ln(f'_1(R))c_4)b-(c_4+c_6)b\ln(b)+c_6b\ln(a),
\\\nonumber\gamma&=&-\frac{2}{\sqrt{f'_1(R)}f''_1(R)b}\left[b((-3c_4-4c_6)\ln(b)
+c_4+c_8+\frac{c_7}{2}\right.\\\nonumber&+&\left.c_6
+c_6\ln(a))(f'_1(R))^{\frac{3}{2}}-c_3f'_1(R)\right],
\end{eqnarray*}
where prime denotes derivative with respect to $R$ and $c_i$
($i=3,4,5,6,7,8$) are arbitrary constants. Inserting these solutions
in Eq.(\ref{21}), we obtain two solutions for $f_1(R)$ as
$f_1(R)=c_9R+c_{10}$ which is similar to the previous case while the
second solution increases the complexity of the system. To avoid
this situation, we consider $f_1(R)=f_0R^n,~(n\neq0,~1)$ which
yields
\begin{eqnarray*}
\alpha&=&ac_{11},\quad\beta=bc_{12},\quad\gamma=\frac{(c_{11}+2c_{12})R}{1-n},\quad
f_2(T)=\frac{T}{3}+c_{13},\\\nonumber
p&=&\frac{1}{12nc_{13}}\left[R^{1-n}b\rho,_b-Rc_{13}
-6R^{1-n}c_{13}+2R^{1-n}\rho+6nc_{13}R\right],
\\\nonumber\rho&=&3f_0R^n+3c_{13}-\frac{(c_{11}a\rho,_a
+c_{12}b\rho,_b)}{(c_{11}+2c_{12})}.
\end{eqnarray*}

These solutions satisfy (\ref{13})-(\ref{23}) for $n=2$ which
implies that $f_1(R)=f_0R^2$ and hence this quadratic curvature term
describes an indirect non-minimal coupling of the matter components
with geometry. Thus the matter contents and model of $f(R,T)$
gravity turn out to be
\begin{eqnarray*}
\rho&=&3f_0R^2+3c_{13}+\frac{a^{-1
+\frac{c_{12}}{c_{11}}}b}{2},\quad
p=\frac{1}{24c_{13}}\left[\frac{3a^{-1+\frac{c_{12}}{c_{11}}}
bR^{-1}}{2}+12c_{13}R\right],
\\\nonumber f(R,T)&=&f_0R^2+\frac{T}{3}+c_{13},\quad
T=3p-\rho.
\end{eqnarray*}
In this case, the constructed $f(R,T)$ model is found to be viable
as it preserves stability conditions (\ref{R}). The corresponding
symmetry generator takes the form
\begin{equation*}
K=ac_{11}\frac{\partial}{\partial a}+bc_{12}\frac{\partial}{\partial
b}-R(c_{11}+2c_{12})\frac{\partial}{\partial R}.
\end{equation*}
This generator yields scaling symmetry with the following conserved
factors
\begin{equation*}
\Sigma_1=4ab^2\dot{R}f_0-4b^2\dot{a}f_0R,\quad\Sigma_2=-24ab\dot{b}f_0R-8ab^2\dot{R}f_0,
\end{equation*}
where $\Sigma_1$ and $\Sigma_2$ are conserved quantities
corresponding to $c_{11}$ and $c_{12}$, respectively.

To reduce the complex nature of the system, we consider $\phi:
(a,b,R)\rightarrow(u,v,z)$ implying that $\phi_Kdu=0$, $\phi_Kdv=0$
and $\phi_Kdz=1$. In this case, we choose $z$ as cyclic variable
which gives
\begin{equation}\nonumber
u=A_0a^{\frac{c_{11}+2c_{12}}{c_{11}}}R,\quad
v=A_1b^{\frac{c_{11}+2c_{12}}{c_{12}}}R,\quad
z=-\frac{1}{c_{11}+2c_{12}}\ln R,
\end{equation}
where $A_0$ and $A_1$ denote integration constants. The
corresponding inverse point transformation yields
\begin{eqnarray}\nonumber
a&=&u^{\frac{c_{11}}{c_{11}+2c_{12}}}e^{c_{11}z},\quad
b=v^{\frac{c_{12}}{c_{11}+2c_{12}}}e^{c_{12}z},\quad
R=e^{c_{11}+2c_{12}z}.
\end{eqnarray}
For these solutions, the Lagrangian (\ref{5}) takes the form
\begin{eqnarray*}\nonumber
\mathcal{L}&=&\frac{1}{(c_{11}+2c_{12})^2}\left(24f_0\dot{z}^2
v^{\frac{2c_{12}}{c_{11}+2c_{12}}}c_{11}^3
u^{\frac{c_{11}}{c_{11}+2c_{12}}}c_{12}
+60f_0\dot{z}^2v^{\frac{2c_{12}}{c_{11}+2c_{12}}}
u^{\frac{c_{11}}{c_{11}+2c_{12}}}\right.\\\nonumber&\times&\left.c_{11}^2c_{12}^2+80f_0\dot{z}^2
v^{\frac{2c_{12}}{c_{11}+2c_{12}}}c_{11}c_{12}^3u^{\frac{c_{11}}{c_{11}+2c_{12}}}
+16f_0\dot{v}\dot{z}u^{\frac{c_{11}}{c_{11}+2c_{12}}}
c_{12}^3v^{-\frac{c_{11}}{c_{11}+2c_{12}}} \right.\\\nonumber
&+&\left.4f_0\dot{u}\dot{z}v^{\frac{2c_{12}}{c_{11}+2c_{12}}}c_{11}^3u^{-\frac{2c_{12}}{c_{11}+2c_{12}}}
-8f_0\dot{u}\dot{v}c_{12}c_{11} v^{-\frac{c_{11}}{c_{11}+2c_{12}}}
u^{-\frac{2c_{12}}{c_{11}+2c_{12}}}
+8f_0\dot{u}\dot{z}\right.\\\nonumber&\times&\left.v^{\frac{2c_{12}}{c_{11}+2c_{12}}}
u^{-\frac{2c_{12}}{c_{11}+2c_{12}}}c_{12}c_{11}^2
+8f_0\dot{v}\dot{z}u^{\frac{c_{11}}{c_{11}+2c_{12}}}
v^{-\frac{c_{11}}{c_{11}+2c_{12}}}c_{12}^2c_{11}-\left(u^{\frac{c_{11}}{c_{11}+2c_{12}}}e^{c_{11}
z}\right)^{\frac{c_{12}}{c_{11}}}\right.\\\nonumber&\times&\left.v^{\frac{3c_{12}}{c_{11}+2c_{12}}}
e^{3c_{12}z}c_{11}^2-4\left(u^{\frac{c_{11}}{c_{11}+2c_{12}}}e^{c_{11}
z}\right)^{\frac{c_{12}}{c_{11}}}v^{\frac{3c_{12}}{c_{11}+2c_{12}}}
e^{3c_{12}z}c_{12}^2-4f_0\dot{v}^2c_{12}^2\right.\\\nonumber&\times&\left.u^{\frac{c_{11}}{c_{11}+2c_{12}}}v^{-\frac{2
(c_{12}+c_{11})}{c_{11}+2c_{12}}}+48f_0\dot{z}^2
u^{\frac{c_{11}}{c_{11}+2c_{12}}}v^{\frac{2
c_{12}}{c_{11}+2c_{12}}}c_{12}^4+4v^{\frac{2c_{12}}{c_{11}+2c_{12}}}
u^{\frac{c_{11}}{c_{11}+2c_{12}}}\right.\\\nonumber&\times&\left.f_0
c_{11}^4\dot{z}^2-4\left(u^{\frac{c_{11}}{c_{11}+2c_{12}}}e^{c_{11}
z}\right)^{\frac{c_{12}}{c_{11}}}v^{\frac{3c_{12}}{c_{11}+2c_{12}}}
e^{3c_{12}z}c_{11}c_{12}\right).
\end{eqnarray*}
Here, the Lagrangian again depends on the cyclic variable $z$.
Consequently, this approach does not provide a successive way to
evaluate exact solution of the anisotropic universe model in this
case.

\section{Noether Gauge Symmetry}

In this section, we determine Noether gauge symmetry of homogeneous
and isotropic as well as anisotropic universe for
$f(R,T)=f_0R^n+h(T)$ model.

\subsection{Flat FRW Universe Model}

We first consider flat FRW metric given by
\begin{equation}\label{34}
ds^2=-dt^2+a^2(t)(dx^2+dy^2+dz^2),
\end{equation}
where the scale factor $a$ describes expansion in $x,~y$ and
$z$-directions. For isotropic universe, the Lagrangian depends on
configuration space $(a,~R,~T)$ with tangent space
$(a,~R,~T,~\dot{a},~\dot{R},~\dot{T})$. The metric variation of
action (\ref{1}) with $\mathcal{L}_m=p(a)$ leads to
\begin{eqnarray}\nonumber
&&\mathcal{L}(a,R,T,\dot{a},\dot{R},\dot{T})=a^3[f(R,T)-Rf_R(R,T)
-Tf_T(R,T)+f_T(R,T)\\\nonumber&&\times(3p(a)-\rho(a))+p(a)]
-6(a\dot{a}^2f_R(R,T)+a^2\dot{a}\dot{R}f_{RR}(R,T)
\\\label{35}&&+a^2\dot{a}\dot{T}f_{RT}(R,T)).
\end{eqnarray}
For Noether gauge symmetry, the vector field $K$ with its first
order prolongation is defined as
\begin{eqnarray}\nonumber
K&=&\tau(t,a,R,T)\frac{\partial}{\partial
t}+\alpha(t,a,R,T)\frac{\partial}{\partial
a}+\beta(t,a,R,T)\frac{\partial}{\partial
R}+\gamma(t,a,R,T)\frac{\partial}{\partial T},\\\nonumber
K^{[1]}&=&\tau\frac{\partial}{\partial
t}+\alpha\frac{\partial}{\partial a}+\beta\frac{\partial}{\partial
R}+\gamma\frac{\partial}{\partial
T}+\dot{\alpha}\frac{\partial}{\partial
\dot{a}}+\dot{\beta}\frac{\partial}{\partial
\dot{R}}+\dot{\gamma}\frac{\partial}{\partial \dot{T}},
\end{eqnarray}
where $\tau,~\alpha,~\beta$ and $\gamma$ are unknown coefficients of
vector field to be determined and the time derivatives of these
coefficients are
\begin{eqnarray}\nonumber
\dot{\alpha}=\frac{\partial\alpha}{\partial
t}+\dot{a}\frac{\partial\alpha}{\partial
a}+\dot{R}\frac{\partial\alpha}{\partial
R}+\dot{T}\frac{\partial\alpha}{\partial
T}-\dot{a}\left\{\frac{\partial\tau}{\partial
t}+\dot{a}\frac{\partial\tau}{\partial
a}+\dot{R}\frac{\partial\tau}{\partial R}
+\dot{T}\frac{\partial\tau}{\partial T}\right\},\\
\nonumber \dot{\beta}=\frac{\partial\beta}{\partial
t}+\dot{a}\frac{\partial\beta}{\partial
a}+\dot{R}\frac{\partial\beta}{\partial
R}+\dot{T}\frac{\partial\beta}{\partial
T}-\dot{R}\left\{\frac{\partial\tau}{\partial
t}+\dot{a}\frac{\partial\tau}{\partial
a}+\dot{R}\frac{\partial\tau}{\partial R}
+\dot{T}\frac{\partial\tau}{\partial T}\right\},\\
\nonumber \dot{\gamma}=\frac{\partial\gamma}{\partial
t}+\dot{a}\frac{\partial\gamma}{\partial
a}+\dot{R}\frac{\partial\gamma}{\partial
R}+\dot{T}\frac{\partial\gamma}{\partial
T}-\dot{T}\left\{\frac{\partial\tau}{\partial
t}+\dot{a}\frac{\partial\tau}{\partial
a}+\dot{R}\frac{\partial\tau}{\partial R}
+\dot{T}\frac{\partial\tau}{\partial T}\right\}.
\end{eqnarray}

The existence of Noether gauge symmetry demands
\begin{equation}\label{36}
K^{[1]}\mathcal{L}+(D\tau)\mathcal{L}=DG(t,a,R,T),
\end{equation}
where $G$ represents gauge function and
$D=\partial_t+\dot{a}\partial_a+\dot{R}\partial_R+\dot{T}\partial_T$.
Substituting the values of vector field, its first order
prolongation and corresponding derivatives of coefficients in
Eq.(\ref{36}), we obtain the following system of equations
\begin{eqnarray}\label{37}
&&\tau,_{_a}=0,\quad\tau,_{_R}=0,\quad\tau,_{_T}=0,\quad G,_{_T}=0,
\\\label{41}&&n(n-1)f_0R^{n-2}a^2\alpha,_{_R}=0,\\\label{43}&&
n(n-1)f_0a^2R^{n-2}\alpha,_{_T}=0,\\\label{42}&&
2a\alpha,_{_T}+(n-1)aR^{-1}\beta,_{_T}=0,\\\label{40}&&
6n(n-1)f_0a^2R^{n-2}\alpha,_{_t}=-G,_{_R},\\\label{39}&&
nf_0R^{n-1}[2a\alpha,_{_t}+(n-1)a^2R^{-1}\beta,_{_t}]=-G,_{_a},
\\\label{38}&&\alpha+(n-1)aR^{-1}\beta+2a\alpha,_{_a}
-a\tau,_{_t}+(n-1)a^2R^{-1}\beta,_{_a}=0,
\\\nonumber&&2(n-1)
R^{-1}\alpha+(n-1)(n-2)aR^{-2}\beta+(n-1)aR^{-1}\alpha,_{_a}+2\alpha,_{_R}
-(n-1)\\\label{44}&&\times
aR^{-1}\tau,_{_t}+(n-1)aR^{-1}\beta,_{_R}=0,\\\nonumber&&
\alpha[3a^2\{f_0R^n(1-n)+h(T)-Th(T),_{_T}+h(T),_{_T}(3p-\rho)+p\}
+a^3\{h(T),_{_T}\\\nonumber&&\times(3p,_{_a}-\rho,_{_a})
+p,_{_a}\}]-n(n-1)f_0a^3R^{n-1}\beta +a^3\gamma
h(T),_{_TT}(3p-\rho-T)\\\label{45}&&+a^3\tau,_{_t}\{f_0R^n(1-n)+h(T)
-Th(T),_{_T}+h(T),_{_T}(3p-\rho)+p\}=G,_{_t}.
\end{eqnarray}
Solving the above system, it follows that
\begin{eqnarray*}
\tau&=&\frac{\xi_4t(3\xi_{11}\xi_2-\xi_3\xi_{10})}{\xi_{11}}+\xi_{13},\quad
\alpha=\xi_4(\xi_2a+\xi_3a^{-1}),\\\nonumber\beta&=&\frac{\xi_4
\xi_3(\xi_{10}+\xi_{11}a^{-2})R} {\xi_{11}(1-n)},\quad G=\frac{\xi_1
t}{2},\quad\gamma=0,
\end{eqnarray*}
where $\xi_i$ are arbitrary constants. For these coefficients, the
symmetry generator becomes
\begin{eqnarray*}
K&=&\left(\frac{\xi_4t(3\xi_{11}\xi_2-\xi_3\xi_{10})}{\xi_{11}}+\xi_{13}\right)
\frac{\partial}{\partial t}+\left(\frac{\xi_4
\xi_3(\xi_{10}+\xi_{11}a^{-2})R}{\xi_{11}(1-n)}\right)\frac{\partial}{\partial
R}\\\nonumber&+&\xi_4(\xi_2a+\xi_3a^{-1})\frac{\partial}{\partial
a}.
\end{eqnarray*}
This generator can be split as
\begin{eqnarray}\nonumber
K_1&=&\frac{\partial}{\partial t},\quad
K_2=\left(\frac{t(3\xi_{11}\xi_2-\xi_3\xi_{10})}{\xi_{11}}\right)
\frac{\partial}{\partial t}+\left(\frac{
\xi_3(\xi_{10}+\xi_{11}a^{-2})R}{\xi_{11}(1-n)}\right)\frac{\partial}{\partial
R}\\\nonumber&+&(\xi_2a+\xi_3a^{-1})\frac{\partial}{\partial a},
\end{eqnarray}
where the first generator corresponds to energy conservation. The
corresponding conserved quantities are
\begin{eqnarray}\nonumber
\Sigma_1&=&-\frac{t(3\xi_{11}\xi_2-\xi_3\xi_{10})}{\xi_{11}}
\left[a^3(f_0R^n(1-n)+\epsilon_0-\frac{\rho}{3})-6(a\dot{a}^2
+(n-1)\right.\\\nonumber&\times&\left.a^2\dot{a}\dot{R}R^{-1})nf_0R^{n-1}\right]
+6anf_0R^{n-1}
(2\dot{a}-(n-1)aR^{-1}\dot{R})\left[(\xi_2a+\xi_3a^{-1})
\right.\\\nonumber&-&\left.\frac{t\dot{a}
(3\xi_{11}\xi_2-\xi_3\xi_{10})}{\xi_{11}}\right]
-6n(n-1)f_0a^2R^{n-2}\dot{a}\left[\frac{
\xi_3(\xi_{10}+\xi_{11}a^{-2})R}{\xi_{11}(1-n)}\right.\\\nonumber&+&\left.
\frac{t\dot{R}(3\xi_{11}\xi_2-\xi_3\xi_{10})}{\xi_{11}}\right],\\
\nonumber \Sigma_2&=&-a^3(f_0R^n(1-n)+\epsilon_0-\frac{\rho}{3})-
6(a\dot{a}^2 +2(n-1)a^2\dot{a}\dot{R}R^{-1})nf_0R^{n-1}.
\end{eqnarray}

\subsection{Bianchi I Universe Model}

Here we investigate Noether gauge symmetry for BI universe model. In
this case, the vector field and corresponding first order
prolongation take the form
\begin{eqnarray}\nonumber
K&=&\tau(t,a,b,R,T)\frac{\partial}{\partial
t}+\alpha(t,a,b,R,T)\frac{\partial}{\partial
a}+\beta(t,a,b,R,T)\frac{\partial}{\partial
b}\\\nonumber&+&\gamma(t,a,b,R,T)\frac{\partial}{\partial
R}+\delta(t,a,b,R,T)\frac{\partial}{\partial T},\\\nonumber
K^{[1]}&=&\tau\frac{\partial}{\partial
t}+\alpha\frac{\partial}{\partial a}+\beta\frac{\partial}{\partial
b}+\gamma\frac{\partial}{\partial R}+\delta\frac{\partial}{\partial
T}+\dot{\alpha}\frac{\partial}{\partial
\dot{a}}+\dot{\beta}\frac{\partial}{\partial
\dot{b}}+\dot{\gamma}\frac{\partial}{\partial
\dot{R}}+\dot{\delta}\frac{\partial}{\partial \dot{T}},
\end{eqnarray}
where
\begin{equation*}
\dot{\alpha}=D\alpha-\dot{a}D\tau,\quad\dot{\beta}=D\beta-\dot{b}D\tau,
\quad\dot{\gamma}=D\gamma-\dot{R}D\tau,\quad\dot{\delta}=D\delta-\dot{T}D\tau.
\end{equation*}
Using the above vector field, its prolongation and coefficients
derivatives in the condition of the existence of Noether gauge
symmetry, we formulate the following system of nonlinear partial
differential equations as
\begin{eqnarray}\label{46}
&&\tau,_{_a}=0,\quad\tau,_{_b}=0,\quad\tau,_{_R}=0,\quad\tau,_{_T}=0,\quad
G,_{_T}=0,\\\label{47}&&b\alpha,_{_R}+2a\beta,_{_R}=0,
\\\label{50}&&b\alpha,_{_T}+2a\beta,_{_T}=0,\\\label{48}&&2\beta,_{_a}
+(n-1)bR^{-1}\gamma,_{_a}=0,\\\label{49}&&2\beta,_{_T}
+(n-1)bR^{-1}\gamma,_{_T}=0,\\\label{51}&&b\alpha,_{_T}
+a\beta,_{_T}+(n-1)abR^{-1}\gamma,_{_T}=0,\\\label{52}&&
n(n-1)f_0R^{n-2}[2b^2\alpha,_{_t}+4ab\beta,_{_t}]=-G,_{_R},\\\label{53}&&
nf_0R^{n-1}[4b\beta,_{_t}+2(n-1)b^2R^{-1}\gamma,_{_t}]=-G,_{_a},\\\label{54}&&
nf_0R^{n-1}[4b\alpha,_{_t}+4a\beta,_{_t}+4(n-1)abR^{-1}\gamma,_{_t}]=-G,_{_b},
\\\label{55}&&\alpha+(n-1)aR^{-1}\gamma+2b\alpha,_{_b}+2a\beta,_{_b}
+2(n-1)abR^{-1}\gamma,_{_b}+a\tau,_{_t}=0,\\\nonumber&&2\beta
+2(n-1)bR^{-1}\gamma+2b\alpha,_{_a}+2a\beta,_{_a}+2b\beta,_{_b}+2(n-1)abR^{-1}
\gamma,_{_a}\\\label{56}&&+(n-1)b^2R^{-1}\gamma,_{_b}-2b\tau,_{_t}=0,\\\nonumber&&
2(n-1)R^{-1}\beta+(n-1)(n-2)bR^{-2}\gamma+(n-1)bR^{-1}
\alpha,_{_a}+2\beta,_{_R}\\\label{57}&&+2(n-1)aR^{-1}\beta,_{_a}
+(n-1)bR^{-1}\gamma,_{_R}-(n-1)bR^{-1}\tau,_{_t}=0,\\\nonumber&&
2(n-1)bR^{-1}\alpha+2(n-1)aR^{-1}\beta+2(n-1)(n-2)abR^{-2}\gamma
+2b\alpha,_{_R}\\\nonumber&&+(n-1)b^2R^{-1}\alpha,_{_b}+2(n-1)abR^{-1}
\beta,_{_b} +2a\beta,_{_R}+2(n-1)abR^{-1}\gamma,_{_R}\\\label{58}&&
-2(n-1)abR^{-1}\tau,_{_t}=0,\\\nonumber&&
b^2\alpha[f_0R^n(1-n)+h(T)-T
h(T),_{_T}+h(T),_{_T}(3p-\rho)+p+a\{h(T),_{_T}\\\nonumber&&\times
(3p,_{_a}-\rho,_{_a})+p,_{_a}\}] +\beta[2ab(f_0R^n(1-n)+h(T)-T
h(T),_{_T}+h(T),_{_T}\\\nonumber&&\times
(3p-\rho)+p)+ab^2\{h(T),_{_T}(3p,_{_b}-\rho,_{_b})+p,_{_b}\}]
-n(n-1)f_0ab^2R^{n-1}\gamma\\\nonumber&&+ab^2\delta
h(T),_{_TT}(3p-\rho-T)+ab^2\tau,_{_t}\{f_0R^n(1-n)+h(T) -T
h(T),_{_T}\\\label{59}&&+h(T),_{_T}(3p-\rho)+p\}=G,_{_t}.
\end{eqnarray}

We solve this system of equations
\begin{eqnarray*}
\tau&=&\eta_1,\quad
G=(\eta_2t+\eta_3)\eta_4\eta_5,\quad\alpha=\eta_5\eta_6a,\quad\beta=\eta_5\eta_6b
,\\\nonumber\gamma&=&\frac{\eta_5\eta_6R}{2(1-n)},\quad\delta=0,\quad
\rho=-\frac{3\eta_2\eta_4(\eta_7+\eta_8\ln
a)}{ab^2\eta_6\eta_8},\\\nonumber
p&=&-\frac{1}{2nf_0}[f_0R^n+R^{1-n}\eta_9-Rnf_0],
\\\nonumber
f(R,T)&=&f_0R^n-\frac{1}{6nf_0}[f_0R^n+R^{1-n}\eta_9-Rnf_0]
-\frac{\eta_2\eta_4(\eta_7+\eta_8\ln a)}{ab^2\eta_6\eta_8},
\end{eqnarray*}
where the constants $\eta_i$ are redefined. The solution of these
coefficients lead to
\begin{eqnarray*}
K&=&\eta_1\frac{\partial}{\partial
t}+\eta_5\eta_6a\frac{\partial}{\partial
a}+\eta_5\eta_6b\frac{\partial}{\partial
b}+\frac{\eta_5\eta_6R}{2(1-n)}\frac{\partial}{\partial R}.
\end{eqnarray*}
This generator can be split as
\begin{eqnarray}\nonumber
K_1=\frac{\partial}{\partial t},\quad K_2=a \frac{\partial}{\partial
a}+b\frac{\partial}{\partial
b}+\frac{R}{2(1-n)}\frac{\partial}{\partial R},
\end{eqnarray}
where the first generator yields energy conservation whereas the
second generator provides scaling symmetry. The corresponding
conserved quantities are
\begin{eqnarray}\nonumber
\Sigma_1&=&-ab^2
[(f_0R^n(1-n)+\epsilon_1-\frac{\rho}{3})-nf_0R^{n-1}(2a\dot{b}^2
+(n-1)R^{-1}(2b^2\dot{a}\dot{R}\\\nonumber&+&4ab\dot{b}\dot{R})+4b\dot{a}\dot{b})]
,\\\nonumber \Sigma_2&=&\eta_2t+\eta_3-4b^2\dot{a}nf_0R^{n-1}.
\end{eqnarray}

\section{Final Remarks}

In this paper, we have discussed Noether and Noether gauge
symmetries of BI universe model in $f(R,T)$ gravity. We have
formulated Noether symmetry generators, corresponding conserved
quantities, matter contents ($p,~\rho$) as well as explicit forms of
generic function $f(R,T)$ for BI model via two theoretical models of
$f(R,T)$ gravity, i.e., $R+2\Lambda+h(T)$ and $f_0R^n+h(T)$. We have
also evaluated Noether gauge symmetries and conserved quantities of
homogeneous isotropic as well as anisotropic universe models for
$f_0R^n+h(T)$ model.

For BI universe model, we have found two Noether symmetry generators
for the first model in which the first generator gives scaling
symmetry. We have solved the system by introducing cyclic variable
which lead to exact solution of the scale factors and $f(R,T)$
model. The graphical behavior of scale factors indicate that the
universe undergoes an expansion in $x,~y$ and $z$-directions. To
evaluate exact solution of the anisotropic universe model for the
second symmetry generator, we have constructed Lagrangian in terms
of cyclic variable. The Lagrangian violates the mapping $\phi_Kdz=1$
as it is not independent of cyclic variable $z$. Thus, the symmetry
generator with scaling symmetry yields exact solution of the
anisotropic universe model. We have investigated graphical behavior
of the cosmological parameters, i.e., Hubble and deceleration
parameters for this solution. This indicates an accelerated
expansion of the universe while EoS parameter corresponds to
quintessence phase. The trajectory of $r$ and $s$ parameters
indicates that the constructed $f(R,T)$ model corresponds to
standard $\Lambda$CDM model. For the second model
($f(R,T)=f_1(R)+f_2(T)$) when $f_1(R)=f_0R^n$, the symmetry
generator provides scaling symmetry for $n=2$. This implies that the
scaling symmetry induces an indirect non-minimal quadratic curvature
matter coupling in this gravity.

Finally, we have discussed Noether gauge symmetry and associated
conserved quantities of flat FRW and BI universe models. The time
coefficient of symmetry generator is found to be $t$ dependent for
FRW universe but becomes constant for BI model while gauge function
is non-zero in both cases. The symmetry generator provides energy
conservation for isotropic universe whereas for anisotropic
universe, we have energy conservation along with scaling symmetry.
In the previous work \cite{a4}, we have formulated exact solution
through Noether symmetry approach for LRS BI universe using $f(R)$
power-law model. The cosmological parameters correspond to
accelerated expanding universe while the EoS parameter describes
phantom divide line from quintessence to phantom phase. The Noether
symmetry generator provides scaling symmetry whereas Noether gauge
symmetry yields energy conservation with constant time coefficient
of symmetry generator and gauge term. Here, we have discussed exact
solution via Noether symmetry for BI model. The cosmological
parameters yield consistent results but EoS parameter corresponds to
phantom era. In case of Noether gauge symmetry, we have found time
dependent gauge term and time coefficient of symmetry generator for
flat FRW model but this time coefficient remains constant for BI
model. Thus, the Noether and Noether gauge symmetries yield more
symmetries for non-minimal curvature matter coupling in $f(R,T)$
gravity as compared to $f(R)$ gravity.

\vspace{0.25cm}

{\bf Acknowledgment}

\vspace{0.25cm}

This work has been supported by the \emph{Pakistan Academy of
Sciences Project}.

\end{document}